\documentclass[prl,showpacs,preprintnumbers,amsmath,amssymb,superscriptaddress,twocolumn]{revtex4}
\usepackage{graphics}

\newcommand{\al}{\alpha}
\newcommand{\be}{\beta}
\newcommand{\ga}{\gamma}
\newcommand{\de}{\delta}
\newcommand{\De}{\Delta}

\newcommand{\pd}{\partial}

\newcommand{\dd}{{\rm d}}

\newcommand{\eps}{\epsilon}

\newcommand{\sph}[1]{\Omega^{^{[#1]}}}
\newcommand{\force}{{\bar f}}
\newcommand{\grav}{{\rm G}}

\begin{document}
\title{Regularization of the linearized gravitational self-force for branes}
\author{Richard A. Battye}
\affiliation{Jodrell Bank Observatory, Department of Physics and
Astronomy, University of Manchester, Macclesfield, Cheshire SK11
9DL, UK}
\author{Brandon Carter}
\affiliation{LUTh, Observatoire de Paris--Meudon, 92195 Meudon, France}
\author{Andrew Mennim}
\affiliation{Department of Applied Mathematics and Theoretical
Physics, Centre for Mathematical Sciences, University of Cambridge,
Wilberforce Road, Cambridge CB3 OWA, UK}
\date{\today --- DRAFT}
\preprint{DAMTP-2003-143}

\begin{abstract}
We discuss the linearized, gravitational self-interaction of a 
brane of arbitrary codimension in a spacetime of arbitrary dimension.
We find that in the codimension two case the gravitational self-force is
exactly zero for a Nambu-Goto equation of state,
generalizing a previous result for a string in four
dimensions. For the case of a 3-brane, this picks out the case of a
six-dimensional brane-world model as having special properties which we
discuss. In particular, we see that bare tension on the
brane has no effect locally, suppressing the cosmological
constant problem.
\end{abstract}

\maketitle

The divergent self-force of a charged point particle,
such as the electron, coupled to electromagnetism has been understood
for many years. Its resolution via the inclusion of an ultra-violet (UV)
cut-off, due to the finite radius of the particle, leads to a
renormalization of the particle's mass and a suppression of the pole
singularity at short distances (see, for example, \cite{Jackson})

This problem is not unique and, in fact, similar problems exist for any
distributional source coupled to any kind of field in any spacetime
dimension. An interesting case  is that of a Nambu-Goto string coupled
to linearized gravity.
It has been shown~\cite{BC95,BC98,Damour98} that the self-force,
regularized in the UV by the core width of the string, $\eps$, and in the
infra-red (IR) by the inter-string separation, $\Delta$, is exactly zero due to
the fact that the induced linearized metric perturbation is orthogonal to the
string worldsheet. This result can be shown to be true at all orders
in perturbation theory, in the case of a static string~\cite{gary1}.

A similar result can be deduced when the string in four dimensions is
coupled to an axion field, represented by a 2-form, and a dilaton, as
well as linearized gravity~\cite{CHH,gary2,brandon}.
For a special choice of couplings, which was predicted~\cite{DH89} in the context of ${\cal N}=1$, $D=10$ Supergravity,
one can show that the combined self-interaction is
zero; the dilaton contribution is negative, which cancels the positive
contribution from the axion field.

It should be noted that the UV regularization of the self-field is not
necessary in the codimension one case,  the hypersurface, where the
behaviour at the brane can be dealt with using junction conditions.
The case of gravity can be dealt with exactly, at all orders, using
the conditions often attributed to Israel~\cite{israel} (although see
ref.~\cite{junction}).
Similar lines of argument lead to the junction conditions at a surface
in Maxwell's theory of electromagnetism~\cite{Jackson}.

The extension of these ideas to higher dimensions has become more relevant
recently with the interest that has arisen in brane-world models.
In these models, the matter of the Standard Model of particle physics
is confined to a four-dimensional subspace, or \emph{brane}, of a
higher-dimensional spacetime, often called the \emph{bulk}.
Two types of model have received particular attention: six-dimensional
models with flat, compact extra dimensions, such as the Arkani-Hamed,
Dimopoulos and Dvali (ADD) model~\cite{ADD}, and five-dimensional
models with warped extra dimensions, such as the Randall--Sundrum (RS)
models~\cite{RS}.
These ideas were originally motivated by the notion of D-branes in
M-Theory, and the desire to alleviate the weak hierarchy problem of UV
quantum field theory (QFT). However, much subsequent work has focused on
their gravitational properties. 

Both models can be extended to higher dimensions.
As we have discussed, the five-dimensional case has no UV divergence,
so does not need to be regularized. When the extra dimensions are
compact, the volume of the extra dimensions gives an effective IR
cut-off scale. In the warped case, the curvature lengthscale of the
bulk spacetime fulfils a similar role. In more general cases it is
clear that some physical phenomena must provide either a UV cut-off
(usually the thickness of the brane) or one in the IR (usually the
distance between branes, or the background curvature length scale). In the
codimension two case, one requires both since the self-field is
proportional to $1/r$ and the divergence of the self-energy  is logarithmic.

One intriguing aspect of brane-worlds with two extra dimensions~\cite{6D} is
that the bare tension of the brane, which represents vacuum energy,
does not appear to gravitate from the point of view of an observer on the
brane; its effects only being felt in the bulk as a modification to the
conical deficit angle. This can be thought of as a \emph{self-tuning}
model, suppressing the cosmological constant problem since the large
variations in the vacuum energy expected due to, for example,
cosmological phase transitions would not be experienced
gravitationally by observers on the brane.

As we have described the study of self-interactions finds applications
in a wide range of research areas, from cosmic defects to
superstring and M-theory.  In this \emph{letter} we discuss systematically the
regularization of the gravitational self-force for extended objects
with any codimension more than one, albeit at linearized order. Our
results will also be relevant to the codimension one case, but as we
have already noted, they are not completely necessary there.
We find that, in the case of a Nambu--Goto brane 
the self-force takes a simple form and can be interpreted as
a renormalization of the tension. In the codimension two case, 
this renormalization is exactly zero, extending the result for
cosmic strings~\cite{BC95,BC98,Damour98} to hyper-strings in arbitrary spacetime
dimension.  Our analysis allows for a general configuration of the
brane and for background curvature on a scale greater than the 
effective width of the brane. It is, therefore, an extension of the
self-tuning cosmological constant idea, in that previous work~\cite{6D} 
has considered only  symmetric, exact solutions in specific
background spacetimes.
We will consider the analogue of refs.~\cite{CHH,gary2,brandon}, which
includes the effect of a dilaton and an antisymmetric 
form field, in a more detailed forthcoming paper.

We will consider a $p$-brane, with $(p+1)$-dimensional worldsheet, in an
$n$-dimensional spacetime. 
The position of the brane will be given in terms of the spacetime
coordinates $x^\mu$ by $x^\mu=X^\mu\{\sigma^a\}$,
where $\sigma^a$ are internal worldsheet coordinates.  The induced
metric on the brane is then given by
$\ga_{ab}=g_{\mu\nu}\pd_aX^\mu\pd_bX^\nu$,
and the background energy-momentum tensor, $\widehat T^{\mu\nu}$,
due to that supported on the
worldsheet, $\overline T^{\mu\nu}$, is
\begin{equation}
\widehat{T}^{\mu\nu}\{x\}=\frac{1}{\sqrt{-g}}\int\overline{T}^{\mu\nu}
\de^{(n)}\Big\{x-X\big\{\sigma^a\big\}\Big\}\sqrt{-\gamma}\,\dd^{p+1}\sigma\,.
\end{equation}

The first fundamental tensor of the brane and its orthogonal
compliment can then be defined as
$\eta^{\mu\nu}=\gamma^{ab}\pd_aX^\mu\pd_bX^\nu$ and
$\perp_{\mu\nu}=g_{\mu\nu}-\eta_{\mu\nu}$, respectively.
These act as the projection operators tangential and orthogonal the worldsheet.
The second fundamental tensor of the worldsheet and the extrinsic
curvature vector are defined as
\begin{equation}
K_{\mu\nu}{}^\rho=\eta_\mu{}^\al\eta_{\nu\be}\nabla_\al\eta^{\be\rho}\,,
\qquad K^\rho=g^{\mu\nu} K_{\mu\nu}{}^\rho\,.
\end{equation}
This formulation in terms of background tensorial quantities has the
advantage of avoiding the complications of then internal indices, $\sigma^a$.
In the case of a codimension one brane, $\perp_{\mu\nu}=n_{\mu}n_{\nu}$,
$K_{\mu\nu}=K_{\mu\nu}{}^\rho n_\rho$ and $K^{\rho}=Kn^{\rho}$ where $n_\rho$
is the unit normal covector to the brane, and $K_{\mu\nu}$, $K$ are
the more familiar extrinsic curvature pseudotensor and scalar respectively. 
Note that we use the sign conventions of~\cite{MTW}, whereas, some
authors define the extrinsic curvature with the opposite sign.

We will perform our regularization calculation in a flat background
spacetime, but it will also be valid in the case where the background
is curved so long as the associated curvature scale is larger than the
brane thickness.
In the case of an Anti-de-Sitter (AdS) background this would require
that the AdS length scale, $l\gg\eps$.
Moreover, we also demand that the curvature scale of the brane be much
larger than the brane thickness, i.e., $\sqrt{K^{\rho}K_{\rho}}\gg\eps$.
This condition would not hold, for example, at a cusp in the brane worldsheet.

We consider a perturbation of the metric $g_{\mu\nu}\rightarrow
g_{\mu\nu}+h_{\mu\nu}$ with $|h_{\mu\nu}| \ll 1$, which, in an $n$-dimensional Minkowski
spacetime, will satisfy the linearized Einstein equation
\begin{equation}
\square h_{\mu\nu}=-2(n-2)\sph{n-2}\grav
\left(\widehat{T}_{\mu\nu}-\frac{1}{n-2}\widehat{T}g_{\mu\nu}\right)\,,
\end{equation}
where $\square=\nabla_\rho\nabla^\rho$ is the wave operator
defined by the unperturbed metric, $\grav=M^{2-n}$,
is the gravitational coupling
constant appropriate to the background spacetime and $\sph{n}$ is the
area of a unit $n$-sphere.

In order to perform the regularization, we make the split of the metric perturbation
$h_{\mu\nu}=\hat{h}_{\mu\nu}+\tilde{h}_{\mu\nu}$, where
$\hat{h}_{\mu\nu}$ is the singular contribution from the string and
$\tilde{h}_{\mu\nu}$ is the (finite) remainder due to radiation
backreaction and  external effects.
Defining the standard Green's function ${\cal G}\big\{x,X\{\sigma^a\}\big\}$ 
for the wave operator in Minkowski space, we find that 
\begin{equation}
h_{\mu\nu}\{x\}=\beta\int\left(\overline{T}_{\mu\nu}-
\frac{1}{n-2}\overline{T}g_{\mu\nu}\right){\cal G}\big\{x,X\{\sigma^{a}\}\big\}d^{p+1}\sigma\,,
\end{equation}
where $\beta=-2(n-2)\sph{n-2}\grav/\sph{n-1}$.
Using the standard form for the Green's function, this solution can be regularized, both in the UV and the IR, to give 
\begin{equation}
\label{hsoln}
\hat{h}_{\mu\nu}=2\grav \left(\overline{T}_{\mu\nu}-
\frac{1}{n-2}\overline{T}g_{\mu\nu}\right)\,F_{\{\De,\eps\}}\,,
\end{equation}
where we have defined a regularization factor
\begin{equation}
F_{\{\De,\eps\}}=\frac{\sph{n-2}\sph{p}}{\sph{n-1}}
\int^\De_\eps s^{p+2-n}ds\,,
\end{equation}
to describe the dependence on the IR and UV cut-offs. At its simplest
level this represents a hard cut-off in source density for $s<\eps$
and $s>\Delta$, but the cut-offs could easily be thought as effective,
representing the envelope of a solution. 
The factor $F_{\{\De,\eps\}}$ encapsulates all of the dependence on the
internal structure of the string and the effect of spacetime
compactification or curvature on $\hat{h}_{\mu\nu}$. 
The regularization can be justified by a more rigorous calculation
as described in~\cite{BCU02}.

The effective UV cut-off scale, $\eps$, is governed by the internal
structure of the brane.  Except in the codimension one case, the
infinitely thin limit leads to a divergence, meaning that the profile of
the brane is always important. 
The profile of the brane will remove the divergence associated with an
infinitely thin source, thus generating an effective thickness,
which we will assume is the same  at every point on the worldsheet.
For the domain wall case, such as the RS
models~\cite{RS}, there is no UV divergence and the infinitely thin
limit can be used.

The IR cut-off scale, $\De$, is necessary when considering branes of
codimension one or two (domain walls and hyper-strings).
 The IR divergence can be removed in several
ways, each of which will generate an effective value of $\De$.
One such way is the usual Kaluza--Klein approach where one
compactifies the extra dimensions on a circle or torus, the radius or
volume of the internal space giving the effective cut-off scale.
Another possibility is to consider an AdS bulk, as in the RS
model, where the different form of the Green's function will mean that
the corresponding integral does not have an IR divergence.
The AdS lengthscale, $l$, will then give an effective cut-off, $\Delta\sim l$,
allowing us to use the solution (\ref{hsoln}).
Another possibility would be to consider a network of branes where the
inter-brane separation would give an IR cut-off scale, as is usually
 assumed to be the case for $p=1$, $n=4$, i.e., cosmic strings.

One can determine the force acting on the brane by considering the
variation of the brane component to the matter action
\begin{equation}
{\cal S}=\int \overline{\cal L} \sqrt{-\gamma}\,\dd^{p+1}\sigma\,.
\end{equation}
Under the perturbation $g_{\mu\nu} \rightarrow g_{\mu\nu}+h_{\mu\nu}$,
the first order change in the Lagrangian is given by
\begin{equation}
\label{actionrenorm}
\overline{\cal L}_{\rm eff}=\overline{\cal L}+\frac{1}{4}h_{\rho\sigma}\overline{T}^{\rho\sigma}\,,
\end{equation}
an extra factor of $1/2$ being the adjustment required to make sure the contributions are only counted once. 
Varying the action \cite{BC95} shows us that the force on the brane
is given by
\begin{equation}
\label{forcedef}
\force^\mu=\frac{1}{2}\overline{T}^{\nu\rho}\nabla^\mu h_{\nu\rho}-\nabla_\nu
\left(\overline{T}^{\nu\rho}h_\rho{}^\mu+\overline{C}^{\mu\nu\rho\sigma}
h_{\rho\sigma}\right)\,,
\end{equation}
where the \emph{hyper-Cauchy tensor} is defined by
\begin{equation}
\overline{C}^{\mu\nu\rho\sigma}=\frac{1}{\sqrt{-\gamma}}\frac{\de}{\de
g_{\mu\nu}} \left(\sqrt{-\gamma}\,\overline{T}^{\rho\sigma}\right)\,,
\end{equation}
which is a relativistic version of the Cauchy elasticity tensor.

In order the evaluate the regularized force from (\ref{forcedef}), we
need to compute the regularized version of the gradient $\nabla_{\rho}
h_{\mu\nu}$. We will do this using the formula derived in
ref.~\cite{BCU02} for scalar field $\phi$,
\begin{equation}
\label{gradeq}
\widehat{\nabla_\rho\phi}=\eta_\rho{}^\sigma\nabla_\sigma\hat\phi+
\frac{1}{2}K_\rho\hat\phi\,,
\end{equation}
can applied to each of the component of $h_{\mu\nu}$ since we are considering linearized interactions. This formula applies
when $p>0$ and when the  codimension is greater than one. The factor
of $1/2$ in the second term should be replaced by $(p-1)/2p$ in the
codimension one case. 
The first term in this formula is just the derivative tangent to the
brane, which is all one would have in the case of a straight brane.
The effect of the curvature is in the second term, which accounts for
the change in orientation of the planes normal to the brane when it is
curved within the bulk.

For a Nambu-Goto type brane, the Lagrangian, energy-momentum tensor
and hyper-Cauchy tensor are given by $\overline{\cal L}=-\lambda=-m^{p+1}$, 
$\overline{T}^{\mu\nu}=-\lambda\eta^{\mu\nu}$ and 
\begin{equation}
C^{\mu\nu\rho\sigma}=\lambda\left(\eta^{\mu(\rho}\eta^{\sigma)\nu}-
\frac{1}{2}\eta^{\mu\nu}\eta^{\rho\sigma}\right)\,,
\end{equation}
where $m$ is a fixed mass scale and $\lambda$ is the tension of the brane.
The singular part of the metric perturbation is given by
\begin{equation}
\label{regsoln}
\hat{h}_{\mu\nu}=2\lambda\grav\,F_{\{\De,\eps\}}\,\left(\frac{p+3-n}{n-2}\eta_{\mu\nu}+
\frac{p+1}{n-2}\perp_{\mu\nu}\right)\,,
\end{equation}
which, in the codimension two case where $p=n-3$, will be purely
orthogonal to the worldsheet.

If one regularizes the force (\ref{forcedef}), using the relation
(\ref{gradeq}) for the regularized gradient and the solution
(\ref{regsoln}), one can deduce that the linearized, gravitational
self-force is given by
\begin{equation}
\label{NGforce}
\hat{f}^\mu=\,\frac{(p+1)(p+3-n)}{2(n-2)} \, \lambda^2 \grav F_{\{\De,\eps\}} \,K^\mu\,,
\end{equation}
for a brane of codimension greater than one.  In the codimension one
case, there will be an additional factor of $(p+1)/p$.  The force
is in the direction of the extrinsic curvature vector, $K^\rho$, which
is normal to the brane worldsheet.
This can be interpreted as a renormalization of the tension of the
brane
\begin{equation}
\label{renorm}
\frac{\lambda_{\rm eff}}{\lambda}=1-\,\frac{(p+1)(p+3-n)}{2(n-2)}
\,\lambda\grav F_{\{\De,\eps\}}\,.
\end{equation}
This renormalization represents a correction to the Lagrangian
$\overline{\cal L}$ of the matter supported on the brane, providing a
term which looks like an effective cosmological constant on the
brane.  It is obvious that this force, and hence the renormalization, will
vanish when $p=n-3$, generalizing the result previously derived for
cosmic strings in four-dimensions~\cite{BC98,Damour98}.

One can see this in much more simple terms, if one considers the
action renormalization, that is, if one substitutes (\ref{regsoln}) into
(\ref{actionrenorm}); (\ref{renorm}) can be re-derived very easily by
using the fact that $\eta_{\mu\nu}\eta^{\mu\nu}=p+1$ and one can then see
directly that if $\hat{h}_{\mu\nu}\propto \perp_{\mu\nu}$, as is the
case when $p=n-3$, then not only is the self-force zero, but so is the
action renormalization. 

For a brane-world, $p=3$, and this special case requires $n=6$. This
phenomena has been pointed out recently by several authors~\cite{6D}
who have studied explicit solutions for specific formulations of
six-dimensional brane-worlds. Here, we have generalized this result to
arbitrary, non-static
configurations at linearized order. One can see that for a
general surface energy momentum tensor ${\overline T}^{\mu\nu}$, the
action renormalization is given by  
\begin{equation} 
{\cal L}_{\rm eff}={\cal L}+\frac{1}{2}\grav F_{\{\Delta,\eps\}}
\left({\overline T}^{\mu\nu}{\overline T}_{\mu\nu}-\frac{1}{n-2}
{\overline T}^2\right)\,.   
\end{equation}

Some of these authors have suggested that this could
have implications for the observed gravitational effects of vacuum
energy in these models since  the vacuum energy can be thought of as
being the origin of the bare tension of the brane-world. Effective QFTs
predict that the energy scale associated with the
vacuum could be as large as the Planck mass, $M_{\rm pl}$, making it
discrepant by a factor of the order $10^{120}$ with the upper bounds
from observation.
If $n=6$ and $p=3$ then our calculation has a simple, but elegant resolution to this
problem; the bare tension only gravitates in the direction orthogonal
to the brane-world, that is, $h_{\mu\nu}\propto\,\perp_{\mu\nu}$, the
gravitational self-force is zero and the renormalized action has no
constant contribution. Hence, the effective cosmological constant as
experienced by observers on the brane would be zero, providing a
self-tuning mechanism for the cosmological constant.
This establishes a link between the self-tuning phenomena at
work here and the self-force. 

We should note that there are some obvious problems associated with
this self-tuning mechanism, not least the fact that there is evidence
to suggest that the 
cosmological constant is non-zero~\cite{type1A}. Moreover,
observations of the angular power spectrum of anisotropies in the
cosmic microwave background~\cite{wmap} suggest that inflation is
ultimate their origin. Since both require some kind of acceleration,
albeit from a scalar field in the case of inflation, if this scenario
were correct then this acceleration could not be due to any kind of
brane-based effect since \emph{all} vacuum energy can do is modify the
bulk solution and can have no effects on the brane. 

One resolution of this would be to generate accelerated expansion from bulk
effects.  A bulk cosmological constant will gravitate and give an
effective cosmological constant as seen by a brane-based observer.
Similarly, accelerated expansion could be driven by a bulk scalar
field.  Of course, one still has a cosmological constant problem in
the bulk: how to determine what mechanism fixes this to be small.  Furthermore,
reheating after bulk-induced inflation could be problematic if the
inflaton is not formed from brane-based matter.

To summarize: the main result of this paper is the generalization of
the non-divergence of the self-force of a Nambu-Goto cosmic string to
a corresponding  hyper-string in arbitrary dimensions under the
assumption that the solution is regularized by some physical phenomena
in the UV and the IR. In fact, we have derived a general formula for
the linearized gravitational self-force in arbitrary co-dimension and
the corresponding renormalization of the bare tension. We have pointed
out the links with recent work on attempts to self-tune the
cosmological constant in 6D brane-world models.
An analysis using a similar method  to this but considering the
gravitational interaction of observable matter supported on the brane,
in the usual brane-world limit where it is small compared to the bare
tension, would give a much firmer foundation to these ideas.

\begin{acknowledgements}
RAB is supported by PPARC and AM is supported by Emmanuel College.
\end{acknowledgements}

\end{document}